\documentclass[journal,singleside,final]{IEEEtran}
\usepackage[nocompress]{cite}
\usepackage{amsmath}
\usepackage{amsthm}
\usepackage{amsfonts}
\usepackage{amssymb}
\usepackage[linesnumbered,ruled,vlined]{algorithm2e}
\usepackage{graphicx}
\usepackage{color}
\usepackage[dvipsnames]{xcolor}
\usepackage[hidelinks]{hyperref}
\usepackage[tight]{subfigure}
\usepackage{enumerate}
\usepackage{multirow}
\usepackage{scalefnt}
\definecolor{orange}{rgb}{1,0.5,0}
\definecolor{grn}{rgb}{0.16, 0.5, 0.0}
\makeatletter
\newcommand{\removelatexerror}{\let\@latex@error\@gobble}
\makeatother

\SetKwInput{KwInput}{Input}                
\SetKwInput{KwOutput}{Output}              
\SetKw{Continue}{continue}
\SetKw{Break}{break}

\hypersetup{
    colorlinks=true,
    citecolor=blue,
    linkcolor=black,
    filecolor=magenta,      
    urlcolor=black,
    pdftitle={Overleaf Example},
    pdfpagemode=FullScreen,
    }

\usepackage[acronym,nonumberlist,style=super,toc]{glossaries}

\makeatletter
\newcommand*{\glsplainhyperlink}[2]{%
  \colorlet{currenttext}{.}
  \colorlet{currentlink}{\@linkcolor}
  \hypersetup{linkcolor=currenttext}
  \hyperlink{#1}{#2}%
  \hypersetup{linkcolor=currentlink}
}
\let\@glslink\glsplainhyperlink
\makeatother

\newacronym{5g}{5G}{fifith generation}
\newacronym{6g}{6G}{sixth generation}
\newacronym{qpsk}{QPSK}{quadrature phase shift keying}

\newacronym{awgn}{AWGN}{additive white Gaussian noise}

\newacronym{bs}{BS}{base station}
\newacronym{ber}{BER}{bit error rate}
\newacronym{bpsk}{BPSK}{binary phase shift keying}

\newacronym{csi}{CSI}{channel state information}
\newacronym{cdf}{CDF}{cumulative  distribution  function}

\newacronym{dl}{DL}{downlink}

\newacronym{iid}{i.i.d}{independent and identically distributed}
\newacronym{iot}{IoT}{Internet of Things}

\newacronym{jml}{JML}{joint multiuser maximum likelihood}

\newacronym{lhs}{LHS}{left-hand side}

\newacronym{map}{MAP}{maximum a posteriori}
\newacronym{ml}{ML}{maximum likelihood}

\newacronym{noma}{NOMA}{non-orthogonal multiple access}

\newacronym{pdf}{PDF}{probability density function}

\newacronym{snr}{SNR}{signal-to-noise ratio}
\newacronym{sic}{SIC}{successive interference cancellation}
\newacronym{sc}{SC}{superposition coding}

\renewenvironment{thebibliography}[1]{
  \begin{oldthebibliography}{#1}
    \setlength{\itemsep}{0.01em}
    \setlength{\parskip}{-0.12em}
}
{
  \end{oldthebibliography}
}

\newtheorem{remark}{Remark}

\begin{document}
 \title{\LARGE On the Detection Optimality and Exact BER Analysis of NOMA}
\author{Tasneem~Assaf,~
\IEEEmembership{Member,~IEEE,} Hamad~Yahya,~\IEEEmembership{Member,~IEEE,} and~Arafat~Al-Dweik,~%
\IEEEmembership{Senior Member,~IEEE}
\thanks{The work of Tasneem Assaf and Arafat Al-Dweik was funded by Khalifa University of Science and Technology through the Research Technology and Innovation Grant, under Project ID: KU-INT-RIG-2024-8474000728. The work of Hamad Yahya was funded by Khalifa University of Science and Technology through the Faculty Startup Grant, under Project ID: KU-INT-FSU-2026-8471000026.}
\thanks{Tasneem Assaf, Hamad Yahya, and Arafat Al-Dweik are with Center for Intelligent Computing, Networks and Sensing (ICONS), Department of Computer and Communications Engineering, Khalifa University of Science and Technology, Abu Dhabi, United Arab Emirates (email: \href{mailto:tasneem.assaf@ku.ac.ae}{tasneem.assaf@ku.ac.ae}, \href{mailto:hamad.myahya@ku.ac.ae}{hamad.myahya@ku.ac.ae}, \href{mailto:dweik@fulbrightmail.org}{dweik@fulbrightmail.org})).

Interested readers can find the source code for this article on \url{https://github.com/hmjasmi/opt_noma_detect}.

}}\maketitle
\scalefont{0.97}
\begin{abstract}
\Gls{noma} detection using \gls{sic} and \gls{jml} has been extensively studied, although their presumed optimality for individual-user \gls{ber} minimization is unestablished. Moreover, existing \gls{ber} analyses overlook the \gls{sic}-induced statistical changes. This letter derives optimal per-user \gls{map} detectors, identifies when the near-user detector reduces to conventional \gls{sic}, and proves the equivalence of conventional \gls{sic} and \gls{jml}. An exact average \gls{ber} analysis accounting for \gls{sic}-induced statistical changes is presented. The results show that the proposed detector achieves modest but consistent \gls{ber} gains at low-to-moderate \gls{snr}.
\end{abstract}
\glsresetall
\glsresetall

\markboth{Draft,~Vol.~xx, No.~xx,
OCT.~2026}{ASSAF \MakeLowercase{\textit{et al.}}: On the Optimality and Exact BER Analysis of NOMA with Imperfect SIC and JML Detectors} 

\section{Introduction}
\IEEEPARstart{T}{he} performance and transceiver design of \gls{noma} has been extensively studied in the literature. Several articles present the power-domain \gls{noma} with \gls{sic} as an optimal capacity-achieving scheme from an information-theoretic perspective \cite{Ding2020-CommLett,Ding2025-SPL}. Furthermore, several articles analyzed the \gls{ber} of \gls{noma} for various channel models, modulation schemes, number of users and detectors including \gls{sic} and \gls{jml} \cite{Semira2021-WCL,Liu2021-TVT,Assaf2021-SJ,Hamad2024-OJCOM,AlaaEldin2025-OJCOMS,Wang2026-CommLett,Wang2025-CommLett}. Although \gls{sic} is well motivated from an information-theoretic perspective, this viewpoint has often led to adopting \gls{sic} as the default detector for \gls{ber} analyses. However, \gls{ber} is a per-user signal-detection metric. Therefore, the detector that minimizes the individual users' \glspl{ber} should first be derived from a statistical signal-processing perspective using the \gls{map} rule.
In addition, the same statistical treatment is also required for an exact \gls{ber} analysis, especially when \gls{sic} is involved.

Recently, Al-Dweik \textit{et al.} \cite{Aldweik2024-WCL,Aldweik2026-archive} derived the conditional and joint post-\gls{sic} \glspl{pdf} for the channel gain and noise. Such \glspl{pdf} were not considered in any \gls{ber} analysis of \gls{noma} systems with \gls{sic}. Motivated by these observations, this letter bridges the gap by: 1) Deriving optimal detectors for the \gls{noma} users to minimize the individual user \gls{ber}. We show that neither the \gls{jml} nor the \gls{sic} is optimal for the near user. 2) The condition under which the optimal detector reduces to the conventional \gls{jml}/\gls{sic} detectors is identified, and the equivalence between \gls{jml} and \gls{sic} is proved. 3) For the first time, exact average \gls{ber} analysis is presented over Rayleigh fading channels, considering the statistical impact of the \gls{sic} decision. 

\section{System and Channel Models}\label{sec:Introduct}
The two-user power-domain downlink \gls{noma} is widely used in the literature and is therefore adopted in this work. The far and near users are denoted by $U_{1}$ and $U_{2}$, respectively. The transmitted baseband  \gls{noma} symbol is expressed as $x=\sqrt{\alpha _{1}}\,s_{1}+\sqrt{\alpha _{2}}\,s_{2}$, where $s_k$ is the unit-power modulated information symbol of $U_k$, $\mathbb{E}\{ |s_k|^2 \}=1$, $k\in\{1,2\}$, and $\alpha_k\in\mathbb{R}$ is the corresponding power-allocation coefficient. Without loss of generality, we
assume $0<\alpha _{k}<1$, $\alpha _{1}+\alpha _{2}=1$, and $\alpha
_{1}>\alpha _{2}$. The information symbol $s_{k}$ is \gls{bpsk} modulated,
i.e., $s_{k}\in \mathbb{S}_k=\{-1,+1\}$ with $s_{k}=2a_{k}-1$ and $a_{k}\in \left\{
0,1\right\} $. This modulation scheme is particularly suitable for
resource-constrained devices, such as \gls{iot} and sensing nodes, due to
its low implementation complexity. Consequently, the \gls{noma} symbol $x$
takes four equally likely values, $x\in \mathbb{S}=\left\{
A_{00},A_{01},A_{10},A_{11}\right\} $, where $A_{ij}=\left( 2i-1\right) 
\sqrt{\alpha _{1}}+\left( 2j-1\right) \sqrt{\alpha _{2}}$, $i,j\in
\left\{ 0,1\right\} $, and $\Pr \left( A_{ij}\right) =1/4$,   $\forall
\left\{ i,j\right\} $.

Assuming perfect channel phase compensation, the received signal
 at the $k$th user is given by 
\begin{equation}
y_{k}=\beta _{k}\,x+n_{k}  \label{Eq:Received_phasecomp}
\end{equation}%
where $\beta _{k}$ denotes the Rayleigh fading
amplitude with $\Omega _{k}=\mathbb{E}\left\{ \beta _{k}^{2}\right\}$, and $%
n_{k}\sim \mathcal{N}\left( 0,\sigma _{n}^{2}/2\right) $. The average and instantaneous \glspl{snr} at $U_k$ are defined as
$\bar{\gamma}_{k}=\Omega_k/\sigma_n^2$ and
$\gamma_k=\beta_k^2/\sigma_n^2$, respectively. 

\section{Optimal Detector for the Far-User ($U_1$)} \label{Sec:Optimaldet_FarUSer}
The process to detect $s_1$ at the far-user can be formulated as a binary hypothesis testing given as
\begin{equation}
 y_1=\left\{ 
\begin{array}{ll}
-\beta_1 A_{1j} + n_1 \text{, } & \overline{\mathcal{H}}=\overline{\mathcal{H}}_{0} \\ 
\beta_1 A_{1j}+ n_1  \text{, } & \overline{\mathcal{H}}=\overline{\mathcal{H}}_{1}
\end{array}%
\right.
\end{equation}%
where $\overline{\mathcal{H}}_0$ is the null hypothesis and $\overline{\mathcal{H}}_1$ is the alternative hypothesis of $U_1$. The \gls{ber} of $U_1$ is defined as
\begin{equation}
P_{\mathrm{e},1} =\Pr\left(\, \overline{\mathcal{H}}_{0}|\overline{\mathcal{H}}_{1}\right) \Pr \left( \,\overline{\mathcal{H}}%
_{1}\right) +\Pr \left( \,\overline{\mathcal{H}}_{1}|\overline{\mathcal{H}}_{0}\right) \Pr \left( \,
\overline{\mathcal{H}}_{0}\right)
\end{equation}
where $\Pr\left(\,{\overline{\mathcal{H}}}_{0}|\overline{\mathcal{H}}_{1}\right)$ and $\Pr\left(\,\overline{\mathcal{H}}_{1}|\overline{\mathcal{H}}_{0}\right)$ denote the conditional error probabilities under 
$\overline{\mathcal{H}}_{1}$, and $\overline{\mathcal{H}}_{0}$, respectively, and
$\Pr\left(\,\overline{\mathcal{H}}_{1}\right)$ and
$\Pr\left(\,\overline{\mathcal{H}}_{0}\right)$ denote the corresponding prior probabilities. The \gls{map} detector that minimizes the \gls{ber} of $U_1$ can be formulated as 
\begin{equation}
\widehat{\overline{\mathcal{H}}}=\arg \max_{\overline{\mathcal{H}}\in \{\overline{\mathcal{H}}_0,\overline{\mathcal{H}}_1\}}\Pr\left(\,\overline{\mathcal{H}}|y_1\right).
\end{equation}%
Applying Bayes' theorem, 
\begin{equation}
\Pr\left(\,\overline{\mathcal{H}}|y_1\right)=\frac{f\left(y_1|\overline{\mathcal{H}}\,\right)\Pr\left(\,\overline{\mathcal{H}}\,\right)}{f(y_1)}
\end{equation}%
where $f\left(y_1|\overline{\mathcal{H}}\,\right)$ is the likelihood function, $\Pr\left(\,\overline{\mathcal{H}}\,\right)$ is the prior probability
of transmitting $\overline{\mathcal{H}}$, and the \gls{pdf} $f(y_1)$ is independent of the decision.
Consequently, the \gls{map} detector becomes
\begin{equation}
\widehat{\overline{\mathcal{H}}}=\arg \max_{{\overline{\mathcal{H}}\in \{\overline{\mathcal{H}}_0,\overline{{\mathcal{H}}}_1\}}}f\left(y_1|\overline{\mathcal{H}}\,\right)\Pr\left(\,\overline{\mathcal{H}}\,\right).
\label{eq:map_6}
\end{equation}
For equal prior probabilities, the detector in \eqref{eq:map_6} can be written as
\begin{equation}
f\left( {y}_{1}|\overline{\mathcal{H}}_{1}\,\right) \overset{\overline{\mathcal{H}}_{1}}{%
\underset{\overline{\mathcal{H}}_{0}}{\gtrless }}f\left( y_{1}|\overline{\mathcal{H}}%
_{0}\,\right) 
\label{eq-MAP-FU}.
\end{equation}
By averaging over the possible hypotheses of $U_2$, $\mathcal{H}\in \{{\mathcal{H}}_0,{\mathcal{H}}_1\}$, we can write $f\left( {y}_{1}|\overline{\mathcal{H}}_{i}\,\right)$ as follows
\begin{align}
f\left( y_1|\overline{\mathcal{H}}_{i}\, \right) & =\sum_{j=0}^{1}f\left( y_1|\overline{\mathcal{H}}_{i} ,\mathcal{H}_j\right)\Pr \left( \mathcal{H}_j  \right)  
   \notag \\
& =\frac{1}{2\sqrt{\pi \sigma_n ^{2}}}\sum_{j=0}^{1}\mathrm{e}^{  -\frac{\left(
y_1-\beta_1 A_{ij}\right) ^{2}}{\sigma_n ^{2}}}\label{eq-GM}
\end{align}%
which is a Gaussian-mixture with equal weights for $\Pr \left( \mathcal{H}_0 \right)=\Pr \left(\mathcal{H}_1 \right)=1/2$. Therefore, \eqref{eq-MAP-FU} can be simplified as
\begin{equation}
    \mathrm{e}^{-\frac{\beta_1^2A_{11}^2}{\sigma_n^2}}\!\sinh\!\left(\frac{2y_1\beta_1 A_{11}}{\sigma_n^2}\right) +  \mathrm{e}^{-\frac{\beta_1^2 A_{10}^2}{\sigma_n^2}}\!\sinh\!\left(\frac{2y_1\beta_1 A_{10}}{\sigma_n^2}\right) \overset{\overline{\mathcal{H}}_{1}}{%
\underset{\overline{\mathcal{H}}_{0}}{\gtrless }} 0.
\label{eq:MAP_U1_00}
\end{equation}
By noting that exponential terms in \eqref{eq:MAP_U1_00} are strictly positive, and that the sign of each $\sinh(\cdot)$ term follows the sign of $y_1$, the decision is determined solely by the sign of $y_1$. Thus, the optimal detector for the far-user can be written as
\begin{equation}
y_1 \overset{\overline{\mathcal{H}}_{1}}{%
\underset{\overline{\mathcal{H}}_{0}}{\gtrless }} 0. \label{eq-ML-FU}
\end{equation}%

\begin{remark} \label{remark1}
 The decision rule in \eqref{eq-ML-FU} is equivalent to the single-user \gls{ml} detector given as%
\begin{equation}
\widehat{s}_{1}=\arg \min_{\tilde{s}_{1}\in \mathbb{S}_{1}}\left\vert y_1-\beta_1 
\sqrt{\alpha_{1}}\tilde{s}_{1}\right\vert ^{2}.\label{eq-SU-MLD}
\end{equation}
where $\tilde{s}_1$ are the trial values of $s_1$.
 \end{remark}
\begin{proof}
Remark \ref{remark1} can be proved by considering the alternative form of \eqref{eq-SU-MLD} given as
\begin{equation}
\left\vert y_1-\beta_1 
\sqrt{\alpha _{1}}\right\vert ^{2}\overset{\overline{\mathcal{H}}_{1}}{%
\underset{\overline{\mathcal{H}}_{0}}{\gtrless }} \left\vert y_1+\beta_1 
\sqrt{\alpha _{1}}\right\vert ^{2}
\end{equation}
which simplifies to \eqref{eq-ML-FU} after expanding and canceling the common terms.
\end{proof}

 \begin{remark} \label{remark2}
The decision rule in \eqref{eq-ML-FU} is equivalent to the \gls{jml} detector that is given by
\begin{equation}
(\widehat{s}_1,\widehat{s}_2)
=
\arg \min_{\tilde{x}\in\mathbb{S}}
\left|
y_k-\beta_k
\tilde{x}
\right|^2 .
\label{eq-JMLD}
\end{equation}
\end{remark}
\begin{proof}
Remark \ref{remark2} can be proved considering the alternative format of \eqref{eq-JMLD} for $U_1$ given as
\begin{multline}
\min\left\{
\left|
y_1-\beta_1A_{11}
\right|^2,\left|
y_1-\beta_1A_{10}
\right|^2\right\}
\overset{\overline{\mathcal{H}}_{1}}{%
\underset{\overline{\mathcal{H}}_{0}}{\gtrless }}\\
\min\left\{
\left|
y_1+\beta_1A_{11}
\right|^2,\left|
y_1+\beta_1A_{10}
\right|^2\right\}.\label{eq-JMLD-FU}
\end{multline}
Since the two sets in \eqref{eq-JMLD-FU} are symmetric around zero, the decision induced by the \gls{jml} detector reduces to \eqref{eq-ML-FU}.
\end{proof}

\section{Optimal Detector for the Near-User ($U_2$)}
\label{Sec:Optimaldet_NearUSer}
Similar to $U_1$ case, the process to detect $s_2$ at the near-user can be formulated as a binary hypothesis testing given as
\begin{equation}
 y_2=\left\{ 
\begin{array}{ll}
\beta_2 A_{i0} + n_2 \text{, } & \mathcal{H}={\mathcal{H}}_{0} \\ 
\beta_2 A_{i1}+ n_2 \text{, }& \mathcal{H}={\mathcal{H}}_{1}
\end{array}%
\right.
\end{equation}
where ${\mathcal{H}}_0$ is the null hypothesis and ${\mathcal{H}}_1$ is the alternative hypothesis of $U_2$, $A_{i0}=\left\{ 
\begin{array}{l}
A_{00}=-A_{11} \\ 
A_{10}%
\end{array}%
\right. $ and $A_{i1}=\left\{ 
\begin{array}{l}
A_{01}=-A_{10} \\ 
A_{11}%
\end{array}%
\right.$. 
The \gls{map} detector for $U_2$ is similar to that in \eqref{eq-MAP-FU}. Therefore, to derive the \gls{map} for $U_2$ we need to compute $f(y_2 | \mathcal{H}_i)$. By averaging over the possible hypotheses of $U_1$, $\overline{\mathcal{H}}\in \{\overline{\mathcal{H}}_0,\overline{\mathcal{H}}_1\}$, we can write $f\left( {y}_{1}|{\mathcal{H}}_{i}\,\right)$ as follows
\begin{align}
f\left( y_{2}|\mathcal{H}_{i}\right) & =\sum_{i=0}^{1}f\left(
y_{2}|\mathcal{H}_{j}, \overline{\mathcal{H}}_i\right) \Pr \left(\, \overline{\mathcal{H}}%
_{i}\right)   \notag \\
& =\frac{1}{2\sqrt{\pi \sigma _{n}^{2}}}\sum_{i=0}^{1}\mathrm{e}^ { -\frac{%
\left( y_{2}-\beta_{2}A_{ij}\right) ^{2}}{\sigma _{n}^{2}} }
\end{align}%
where $\Pr \left(\, \overline{\mathcal{H}}_0 \right)=\Pr \left(\,\overline{\mathcal{H}}_1 \right)=1/2$. Unlike $U_1$, the Gaussian-mixture for $U_2$ spans both positive and negative sides for $\mathcal{H}_j$.  Therefore, 
\begin{align}
\frac{\sinh \left( \frac{2y_{2}\beta_{2}A_{11}}{\sigma_{n}^{2}}\right) }{%
\sinh \left( \frac{2y_{2}\beta_{2}A_{10}}{\sigma _{n}^{2}}\right) }
&\overset{{\mathcal{H}}_{1}}{%
\underset{{\mathcal{H}}_{0}}{\gtrless }} \frac{\mathrm{e}^{-\frac{\beta _{2}^{2}A_{10}^{2}}{\sigma _{n}^{2}%
} }}{\mathrm{e}^{  -\frac{\beta _{2}^{2}A_{11}^{2}}{\sigma _{n}^{2}}%
 }}.
\end{align}
Consequently, the optimal detector for the near-user can be written as
\begin{equation}
y_{2}-\beta _{2}\sqrt{\alpha _{1}}\,\widehat{s}_{1}\overset{%
\mathcal{H}_{1}}{\underset{\mathcal{H}_{0}}{\gtrless }}\varepsilon (y_{2})  
\label{eq-ML-NU}
\end{equation}%
where $\widehat{s}_1=\mathrm{sgn}(y_2)$, and  
\begin{equation}
\varepsilon (y_{2})\!=-\widehat{s}_1\frac{\sigma _{n}^{2}}{4\beta _{2}\sqrt{\alpha _{2}}}\!\ln \!\!\left( \!\frac{1\!-\!\mathrm{e}^{ \!-\frac{%
4|y_{2}|\beta _{2}A_{11}}{\sigma _{n}^{2}}\! }}{1\!-\!\mathrm{e}%
^{ \!-\frac{4|y_{2}|\beta _{2}A_{10}}{\sigma _{n}^{2}}\! }}%
\!\right) .  \label{eq-NU-thresh}
\end{equation}

\begin{remark}\label{remark3}
The optimal detector for the near-user in \eqref{eq-ML-NU} reduces to the conventional \gls{sic} detector at high \gls{snr}, 
    $\lim_{\sigma_n^2 \to 0} \varepsilon(y_2) = 0 $. Therefore, 
\begin{equation}
y_{2}-\beta_{2}%
\sqrt{\alpha_{1}}\,\widehat{s}_1\overset{%
\mathcal{H}_{1}}{\underset{\mathcal{H}_{0}}{\gtrless }} 0
\label{eq-SIC-NU}
\end{equation}%
 which is the conventional \gls{sic} detector, given by
 \begin{equation}
 \widehat{s}_{2}=\arg \min_{\tilde{s}_{2}\in \mathbb{S}_{2}}\left\vert y_{2}-{%
 \beta}_{2}\sqrt{\alpha _{1}}\,\widehat{s}_{1}-\beta_2\sqrt{\alpha_{2}}\tilde{s}%
 _{2}
 \right\vert ^{2}.  \label{Eq:SIC_Eq}
 \end{equation}
 \end{remark}
 \begin{proof}
    Remark \ref{remark3} can be proved by considering  the alternative form of \eqref{Eq:SIC_Eq} given as
 \begin{equation}
 \left\vert y_2\!-\!\beta_2 
 \sqrt{\alpha _{1}}\,\widehat{s}_1\!-\!\beta_2 
 \sqrt{\alpha _{2}}\right\vert ^{2}\!\overset{%
\mathcal{H}_{1}}{\underset{\mathcal{H}_{0}}{\gtrless }}\!\!
 \left\vert y_2\!-\!\beta_2 
 \sqrt{\alpha_{1}}\,\widehat{s}_1\!+\!\beta_2 
 \sqrt{\alpha_{2}}\right\vert ^{2}
 \end{equation}
 which can be simplified to \eqref{eq-SIC-NU}. 
 \end{proof}

\begin{remark}\label{remark4}
The conventional \gls{sic} detector in \eqref{Eq:SIC_Eq} and \gls{jml} detector for the near-user in \eqref{eq-JMLD} are equivalent.
\end{remark}
\begin{proof}
    Remark \ref{remark4} can be proved considering the three decision region boundaries of the \gls{jml} detector for $\widehat{s}_2$, which can be written as
\begin{equation}
\widehat{s}_{2}=%
\begin{cases}
+1, & y\in \left\{ \left[-\beta_2\sqrt{\alpha_1},0\right] ,\left[ \beta_2\sqrt{\alpha_1},\infty \right] \right\}  \\ 
-1, & y\in \left\{ \left[ -\infty ,-\beta_2\sqrt{\alpha_1} \right] ,\left[ 0,\beta_2\sqrt{\alpha_1}\right] \right\} 
\end{cases}%
.  \label{eq-JMLD-NU}
\end{equation}
Furthermore, considering the two cases for $\widehat{s}_1=\pm1$ and substituting  in \eqref{eq-SIC-NU}, the ranges in \eqref{eq-JMLD-NU} can be found accordingly. To show this, consider the following:\\
\underline{Case 1:} $\widehat{s}_1=+1$ is achieved when $y_2>0$. Therefore, \eqref{eq-SIC-NU} can be written as $y_{2}-\beta _{2}%
\sqrt{\alpha _{1}} \overset{%
\mathcal{H}_{1}}{\underset{\mathcal{H}_{0}}{\gtrless }} 0$. Consequently, 
    $\widehat{s}_2=\left\{ 
\begin{array}{ll}
+1, &  y_2>\beta_2\sqrt{\alpha_1} \\ 
-1, &  0<y_2<\beta_2\sqrt{\alpha_1} \\ 
\end{array}%
\right.$ which represent the first two decisions in \eqref{eq-JMLD-NU}.\\
\underline{Case 2:} $\widehat{s}_1=-1$ is achieved when $y_2<0$. Therefore, \eqref{eq-SIC-NU} can be written as $y_2+\beta_2%
\sqrt{\alpha _{1}}\overset{%
\mathcal{H}_{1}}{\underset{\mathcal{H}_{0}}{\gtrless }} 0$. Consequently,
    $ \widehat{s}_2=\left\{ 
\begin{array}{ll}
+1, &  -\beta_2\sqrt{\alpha_1}<y_2<0 \\ 
-1, &  y_2<-\beta_2\sqrt{\alpha_1} 
\end{array}%
\right.$ which represent the last two decisions in \eqref{eq-JMLD-NU}. 
\end{proof}
\begin{remark} \label{remark5}
The \gls{jml} detector for $U_2$ in \eqref{eq-JMLD} and the derived decision rule in \eqref{eq-ML-NU} are not equivalent. 
\end{remark}
\begin{proof}
    Remark \ref{remark5} can be proved by showing the alternative format of the \gls{jml} detector for the near-user
\begin{multline}
\max\left\{f\left(
y_{2}|\mathcal{H}_{1},\overline{\mathcal{H}}_{1}\right),f\left(
y_{2}|\mathcal{H}_{1},\overline{\mathcal{H}}_{0}\right)\right\}
\overset{%
\mathcal{H}_{1}}{\underset{\mathcal{H}_{0}}{\gtrless }}\\
\max\left\{f\left(
y_{2}|\mathcal{H}_{0},\overline{\mathcal{H}}_{1}\right),f\left(
y_{2}|\mathcal{H}_{0},\overline{\mathcal{H}}_{0}\right)\right\}
\label{eq-JMLD-FU-alt}
\end{multline}
which is different from \eqref{eq-ML-NU}, which can be expressed as
\begin{multline}
    \sum\nolimits_{i=0}^{1}f\left(
y_{2}|\mathcal{H}_{1}, \overline{\mathcal{H}}_i\right) \Pr \left(\, \overline{\mathcal{H}}%
_{i}\right)\overset{%
\mathcal{H}_{1}}{\underset{\mathcal{H}_{0}}{\gtrless }} \\
\sum\nolimits_{i=0}^{1}f\left(
y_{2}|\mathcal{H}_{0}, \overline{\mathcal{H}}_i\right) \Pr \left(\, \overline{\mathcal{H}}%
_{i}\right).
\end{multline}
The former minimizes the \gls{noma} symbol error rate, while the latter minimizes the near-user \gls{ber}.
\end{proof}

\gls{bpsk} is considered for analytical tractability and to satisfy space limitations.  The analysis can be applied to \gls{qpsk} by treating the in-phase and quadrature components of the \gls{noma} signal as two orthogonal \gls{bpsk} signals. 

\section{Exact BER Analysis for the Near-User ($U_2)$}
\label{Sec:BER_Optimal_NearUser}
While $U_1$ \gls{ber} analysis is omitted since the exact analysis is available in \cite[Eq. (28)]{Hamad2024-OJCOM}, this section presents the exact \gls{ber} for $U_2$ considering the optimal detector in \eqref{eq-ML-NU}. The analysis explicitly accounts for the statistical impact of the \gls{sic} decision. After detecting $s_{1}$, two mutually exclusive events may occur, a successful \gls{sic} event $\mathcal{S}\triangleq \{\widehat{s}_{1}=s_{1}\}$, and an unsuccessful \gls{sic} event $\mathcal{F}\triangleq \{\widehat{s}_{1}\neq s_{1}\}$ \cite{Aldweik2026-archive}. For brevity, the user index is dropped from the fading coefficient, noise sample, and received signal.
Since \gls{sic} is a decision-directed operation, conditioning on $\mathcal{S}$ or $\mathcal{F}$ changes the statistical behavior of both the fading and
the noise. Specifically, $\beta |\mathcal{S}\rightarrow \beta _{\mathcal{S}},\, n|\mathcal{S}\rightarrow n_\mathcal{S}$, and $\beta |\mathcal{F}\rightarrow
\beta _{\mathcal{F}},\, n|\mathcal{F}\rightarrow n_\mathcal{F}.$ Additionally, although $\beta$ and $n$ are independent, the condition on the \gls{sic} outcome transforms them to become jointly dependent \cite{Aldweik2026-archive}. Therefore, the post-\gls{sic} fading and noise should not be treated as independent variables with their original marginal distributions, a widely considered assumption in the literature. Instead, the joint post-\gls{sic} \glspl{pdf} must be used.

\subsection{The Optimal Boundary}
By defining $f\left(
y|A_{ij}\right)\triangleq f\left(
y|\mathcal{H}_{j}, \overline{\mathcal{H}}_i\right)$, the optimal decision boundary in \eqref{eq-ML-NU} can be defined as the crossing point where $f(y|A_{10})=f(y|A_{11})$ or $f(y|A_{00})=f(y|A_{01})$. Owing to the symmetry around the origin, these crossings occur in symmetric pairs, hence, characterizing the positive crossing suffices. Let $\tau_{\beta}$ denote the positive crossing conditioned on $\beta$, $\beta > 0$. The boundary $\tau_{\beta}$ is determined by the equality condition of the optimal detector. In particular, for a received sample $y > 0$, this condition can be expressed as 
\begin{equation}
\tau _{\beta }=-\frac{\sigma _{n}^{2}}{4\beta \sqrt{\alpha_2}}\ln
\left( \frac{1-\mathrm{e}^{ -\frac{4\tau_\beta\beta A_{11}}{\sigma _{n}^{2}}%
} }{1-\mathrm{e}^{  -\frac{4\tau_\beta\beta A_{10}}{\sigma _{n}^{2}}%
} }\right) +\beta \sqrt{\alpha _{1}}.  \label{eq:tau_original_condition}
\end{equation}%
It should be noted that the statistical behavior of $\tau_\beta$ changes according to the  post-\gls{sic} fading marginal distributions, i.e., $\tau_\beta|\mathcal{F} \rightarrow \tau_{\beta_\mathcal{F}}$ and $\tau_\beta|\mathcal{S} \rightarrow \tau_{\beta_\mathcal{S}}$.  Furthermore, the optimal decision regions of the near-user are given as
\begin{equation}
\widehat{s}_{2}=%
\begin{cases}
+1, & y\in \left\{ \left[ -\tau _{\beta },0\right] ,\left[ \tau _{\beta
},\infty \right] \right\}  \\ 
-1, & y\in \left\{ \left[ -\infty ,-\tau _{\beta }\right] ,\left[ 0,\tau
_{\beta }\right] \right\} 
\end{cases}%
.  \label{eq:Opt_Decision_Regions}
\end{equation}
To obtain a compact representation of $\tau _{\beta }$, define $\varkappa _{\beta }\triangleq \mathrm{e}^{ -\frac{4\tau _{\beta }\beta \sqrt{\alpha _{2}}}{\sigma _{n}^{2}}} ,\, 0<\varkappa _{\beta }\leq 1$, $r\triangleq \sqrt{\frac{\alpha _{1}}{\alpha _{2}}},\, \Lambda
_{\beta }\triangleq \mathrm{e}^{\frac{4\beta ^{2}\sqrt{\alpha _{1}\alpha _{2}%
}}{\sigma _{n}^{2}}}$. Since $\frac{A_{11}}{\sqrt{\alpha _{2}}}=r+1$
and $\frac{A_{10}}{\sqrt{\alpha _{2}}}=r-1,$ then $\mathrm{e}^{ -\frac{4\tau
_{\beta }\beta A_{11}}{\sigma _{n}^{2}}} =\varkappa _{\beta }^{r+1}$
and $\mathrm{e}^{ -\frac{4\tau _{\beta }\beta A_{10}}{\sigma _{n}^{2}}}
=\varkappa _{\beta }^{r-1}.$ Substituting these identities into %
\eqref{eq:tau_original_condition} gives 
\begin{equation}
\frac{1-\varkappa _{\beta }^{r+1}}{1-\varkappa _{\beta }^{r-1}}=\Lambda
_{\beta }\varkappa _{\beta }.  \label{eq:kappa_ratio}
\end{equation}%
Consequently, $\varkappa _{\beta }$ is obtained from 
\begin{equation}
\varkappa _{\beta }^{r+1}-\Lambda _{\beta }\varkappa _{\beta }^{r}+\Lambda
_{\beta }\varkappa _{\beta }-1=0.  \label{eq:kappa_root} 
\end{equation}%
Therefore, ${\tau }_{\beta }$ can be written compactly as
${\tau }_{\beta }{=-\frac{\sigma _{n}^{2}}{4\beta \sqrt{\alpha _{2}}}\ln
\left( \varkappa _{\beta }^{\star }\right) } $,
where $\varkappa _{\beta }^{\star }\in (0,1]$ is the root of %
\eqref{eq:kappa_root}. A non-zero positive boundary exists only when $\Lambda
_{\beta }>\frac{A_{11}}{A_{10}}.$ Otherwise, $\tau_\beta=0$. 
For arbitrary $\alpha _{1}$ and $\alpha _{2}$, the parameter $r$ is
generally non-integer and may be irrational. Consequently, %
\eqref{eq:kappa_root} is not, in general, a polynomial with fixed finite
degree in $\varkappa _{\beta }$. Hence, a universal closed-form solution for 
$\tau _{\beta }$ does not exist.
\subsection{SIC-Conditioned Prior Probabilities}
The \gls{ber} conditioned on  $\mathcal{E}\in \{\mathcal{S},\mathcal{F}\}$ can be written as 
\begin{equation}
P_{\mathrm{e}|\mathcal{E}}^{\mathrm{opt}}=\sum_{i=0}^{1}\sum_{j=0}^{1}P_{%
\mathrm{e}|A_{ij},\mathcal{E}}^{\mathrm{opt}}\Pr (A_{ij}|\mathcal{E})
\label{eq:Total_Prob_Generic}
\end{equation}%
where $P_{\mathrm{e}|A_{ij},\mathcal{E}}^{\mathrm{opt}}$ is the conditional
error probability given that $A_{ij}$ is transmitted and the \gls{sic}
outcome is $\mathcal{E}$, and $\Pr (A_{ij}|\mathcal{E})$ is the prior probability. Using Bayes' theorem
\begin{equation}
\Pr (A_{ij}|\mathcal{E})=\frac{\Pr (\mathcal{E}|A_{ij})\Pr (A_{ij})}{\Pr (%
\mathcal{E})} . \label{eq:Bayes_Generic}
\end{equation}%

For $d>0$, we define 
\begin{equation}
G(d)\triangleq \Pr (n<\beta d)=\int_{0}^{\infty }\Phi \left( \frac{\sqrt{2}\beta d}{\sigma _{n}}\right) f_{\beta }(\beta )d\beta   \label{eq:G_def}
\end{equation}
where $\Phi (\cdot )$ is the standard Gaussian \gls{cdf}. For Rayleigh fading,  $G(d)=\frac{1+\mathcal{M}(d)}{2}$ where  $\mathcal{M}(d)\triangleq \sqrt{\frac{%
d^{2}\bar{\gamma}}{d^{2}\bar{\gamma}+{1}}}$. 

The conditional \gls{sic}-success probabilities in \eqref{eq:Bayes_Generic}  can be
expressed as 
\begin{equation}
    p_{11} \triangleq \Pr (\mathcal{S}|A_{00})=\Pr (\mathcal{S}%
|A_{11})=G(A_{11})  \label{eq:p11_def}
\end{equation}
\begin{equation}
    p_{10}\triangleq \Pr (\mathcal{S}|A_{01})=\Pr (\mathcal{S}%
|A_{10})=G(A_{10}).  \label{eq:p10_def}
\end{equation}
The corresponding failure probabilities are $q_{11}=1-p_{11}$ and $%
q_{10}=1-p_{10}$. Therefore, the total success and failure probabilities can be computed as 
\begin{equation}
\Pr (\mathcal{S})=\frac{p_{11}+p_{10}}{2}\text{ \ \ \ and \ \ }\Pr (\mathcal{%
F})=\frac{q_{11}+q_{10}}{2}.  \label{eq:Pr_SF}
\end{equation}%
Finally, the conditional priors in \eqref{eq:Bayes_Generic} are 
\begin{align}
\Pr (A_{00}|\mathcal{S})& =\frac{0.5p_{11}}{p_{11}+p_{10}}\text{, }\Pr
(A_{01}|\mathcal{S})=\frac{0.5p_{10}}{p_{11}+p_{10}}  \notag \\
\Pr (A_{00}|\mathcal{F})& =\frac{0.5q_{11}}{q_{11}+q_{10}}\text{, }\Pr
(A_{01}|\mathcal{F})=\frac{0.5q_{10}}{q_{11}+q_{10}}.
\label{eq:Conditional_Priors}
\end{align}
By symmetry, $\Pr (A_{00}|\mathcal{E})=\Pr (A_{11}|\mathcal{E})$ and $\Pr (A_{01}|\mathcal{E})=\Pr (A_{10}|\mathcal{E})$. It is important to highlight that an extensive body of the literature assumes that the post-\gls{sic} fading and noise statistics remain unchanged. Hence, $\Pr(A_{ij}|\mathcal{E})$ in \eqref{eq:Conditional_Priors} is considered equally likely, i.e. $\Pr(A_{ij}|\mathcal{E})=\frac{1}{4}$, which is  inaccurate \cite[Eqs. (55), (58)]{Hamad2024-OJCOM}.

\subsection{BER Conditioned on Successful SIC}
To evaluate the \gls{ber} in \eqref{eq:Total_Prob_Generic}, the error probability must be computed for each transmitted \gls{noma} symbol while accounting for the corresponding \gls{sic} outcome. Consider first the case in which $A_{00}$ is transmitted, so that $s_2 = -1$. Conditioned on successful \gls{sic}, the resulting observation is $y_{\mathcal{S}} = -\beta_{\mathcal{S}}A_{11} + n_{\mathcal{S}}$. A detection error occurs when the receiver decides $\widehat{s}_2 = +1$, which, according to \eqref{eq:Opt_Decision_Regions}, corresponds to the interval $-\tau{\beta_{\mathcal{S}}} < y_{\mathcal{S}} < 0$, or equivalently, $\beta_{\mathcal{S}}A_{11}-\tau_{\beta_{\mathcal{S}}}
<n_{\mathcal{S}}<
\beta_{\mathcal{S}}A_{11}$.  Thus,
\begin{equation}
P_{\mathrm{e}|A_{00},\mathcal{S}}^{\mathrm{opt}}=\!\!\int_{0}^{\infty
}\!\!\!\!\int_{\beta _{\mathcal{S}}A_{11}-\tau _{\beta _{\mathcal{S}}}}^{\beta _{%
\mathcal{S}}A_{11}}\!\!\!\!f(\beta _{%
\mathcal{S}},n_{\mathcal{S}}|A_{00})dn_{\mathcal{S}}\,d\beta _{\mathcal{S}} 
=\frac{I_{11}^{\mathcal{S}}}{p_{11}}  \label{eq:Pe_A00_S}
\end{equation}
where  the joint \gls{pdf} $f(\beta _{%
\mathcal{E}},n_{\mathcal{E}}|A_{ij}),$ $\forall \left\{ i,j,\mathcal{E}\right\} $
is defined in \cite{Aldweik2026-archive} and  $I_{11}^{\mathcal{S}}\triangleq\int_{0}^{\infty
}\left[ \Phi \left( \frac{\sqrt{2}\beta _{\mathcal{S}}A_{11}}{\sigma _{n}}\right)
-\Phi \left( \frac{\sqrt{2}(\beta _{\mathcal{S}}A_{11}-\tau _{\beta _{\mathcal{S}}})}{%
\sigma _{n}}\right) \right]f_{\beta }(\beta _{\mathcal{S}})d\beta _{%
\mathcal{S}}.$ By symmetry, $P_{\mathrm{e}|A_{11},\mathcal{E}}^{\mathrm{opt}%
}=P_{\mathrm{e}|A_{00},\mathcal{E}}^{\mathrm{opt}}.$

For $A_{01}=-A_{10}$, $s_{2}=+1$ and $y_{\mathcal{S}}=-\beta _{\mathcal{S}%
}A_{10}+n_{\mathcal{S}}.$ Given successful \gls{sic}, an error occurs when
the detector decides $\widehat{s}_{2}=-1$, i.e., $y_{\mathcal{S}%
}<-\tau _{{\beta }_{\mathcal{S}}}$ or $n_{\mathcal{S}}<\beta _{%
\mathcal{S}}A_{10}-\tau _{{\beta }_{\mathcal{S}}}.$ Hence, 
\begin{equation}
P_{\mathrm{e}|A_{01},\mathcal{S}}^{\mathrm{opt}}=\!\!\int_{0}^{\infty
}\!\!\!\!\int_{-\infty }^{\beta _{\mathcal{S}}A_{10}-\tau _{{\beta }_{%
\mathcal{S}}}}\!\!\!\!f(\beta _{%
\mathcal{S}},n_{\mathcal{S}}|A_{01})dn_{\mathcal{S}}\,d\beta _{\mathcal{S}}  =\frac{I_{10}^{\mathcal{S}}}{p_{10}}  \label{eq:Pe_A01_S}
\end{equation}%
where $I_{10}^{\mathcal{S}}\hspace{-1mm}\triangleq \hspace{-1mm}\int_{0}^{\infty }\Phi \left( \frac{\sqrt{2}(
\beta _{\mathcal{S}}A_{10}-\tau _{{\beta }_{\mathcal{S}}})}{\sigma
_{n}}\right)\hspace{-1mm} f_{\beta }(\beta _{\mathcal{S}})d\beta _{\mathcal{S}}.$ By
symmetry, $P_{\mathrm{e}|A_{10},\mathcal{E}}^{\mathrm{opt}}=P_{\mathrm{e}%
|A_{01},\mathcal{E}}^{\mathrm{opt}}.$
Substituting \eqref{eq:Conditional_Priors}, \eqref{eq:Pe_A00_S}, and  \eqref{eq:Pe_A01_S} into %
\eqref{eq:Total_Prob_Generic}, the \gls{ber} conditioned on successful \gls{sic} can be expressed as
\begin{equation}
P_{\mathrm{e}|\mathcal{S}}^{\mathrm{opt}}=\frac{I_{11}^{\mathcal{S}}+I_{10}^{%
\mathcal{S}}}{p_{11}+p_{10}}.  \label{eq:Pe_S_final}
\end{equation}
\subsection{BER Conditioned on Unsuccessful SIC}
Starting with $A_{00}$, where  $s_{2}=-1$ and $y_{\mathcal{F}}=-\beta _{\mathcal{F}%
}A_{11}+n_{\mathcal{F}}.$ Given unsuccessful \gls{sic}, $y_{\mathcal{F}}>
0$,  an error occurs when the detector decides $\widehat{s}_{2}=+1$, which
corresponds to $y_{\mathcal{F}}>\tau _{{\beta }_{\mathcal{F}}}$ and 
 $n_{\mathcal{F}}>\beta _{\mathcal{F}}A_{11}+\tau _{{\beta }_{%
\mathcal{F}}}.$ Therefore, 
\begin{equation}
P_{\mathrm{e}|A_{00},\mathcal{F}}^{\mathrm{opt}} \!=\!\!\int_{0}^{\infty
}\!\!\!\!\int_{\beta _{\mathcal{F}}A_{11}+\tau _{{\beta }_{\mathcal{F}%
}}}^{\infty }\!\!\!\!f(\beta _{\mathcal{F}},n_{\mathcal{F}}|A_{00})dn_{\mathcal{F}}d\beta _{%
\mathcal{F}}  \!=\!\frac{I_{11}^{\mathcal{F}}}{q_{11}}  \label{eq:Pe_A00_F}
\end{equation}%
where $I_{11}^{\mathcal{F}}\triangleq \int_{0}^{\infty }\left[ 1-\Phi \left( 
\frac{\sqrt{2}(\beta _{\mathcal{F}}A_{11}+\tau _{{\beta }_{\mathcal{F}})}}{%
\sigma _{n}}\right) \right] f_{\beta }(\beta _{\mathcal{F}})d\beta _{%
\mathcal{F}}.$ 

For $A_{01}$, $s_{2}=+1$ and $y_{\mathcal{F}}=-\beta _{\mathcal{F}%
}A_{10}+n_{\mathcal{F}}.$ Given unsuccessful \gls{sic}, an error occurs when the detector decides $\widehat{s}_{2}=-1$, i.e.,
 $0<y_{\mathcal{F}}<\tau _{{\beta }_{\mathcal{F}}}$ and 
$\beta _{\mathcal{F}}A_{10}<n_{\mathcal{F}%
}<\beta _{\mathcal{F}}A_{10}+\tau_{\beta _{\mathcal{F}}}$. Hence, 
\begin{equation}
P_{\mathrm{e}|A_{01},\mathcal{F}}^{\mathrm{opt}}\!=\!\!\int_{0}^{\infty
}\!\!\!\!\int_{\beta _{\mathcal{F}}A_{10}}^{\beta _{\mathcal{F}}A_{10}+\tau _{{\beta }_{\mathcal{F}}}}\!\!\!\!f(\beta _{\mathcal{F}},n _{{\mathcal{F}}}|A_{01})dn_{\mathcal{F}}\,d\beta _{\mathcal{F}} \!=\!\frac{I_{10}^{\mathcal{F}}}{q_{10}}  \label{eq:Pe_A01_F} 
\end{equation}%
where $I_{10}^{\mathcal{F}}\hspace{-1mm}\triangleq\hspace{-1.5mm}  \int_{0}^{\infty }\hspace{-1.5mm} \left[ \Phi \hspace{-1mm}\left( 
\frac{\sqrt{2}(\beta _{\mathcal{F}}A_{10}+\tau _{{\beta }_{\mathcal{F}}})}{%
\sigma _{n}}\hspace{-1mm} \right)\hspace{-1mm} -\hspace{-1mm}\Phi \left(\hspace{-1mm} \frac{\sqrt{2}\beta _{\mathcal{F}}A_{10}}{\sigma
_{n}}\right)\hspace{-1mm} \right]\hspace{-1.5mm} f_{\beta }(\beta _{\mathcal{F}})d\beta _{\mathcal{F}}.$

Substituting  \eqref{eq:Conditional_Priors}, \eqref{eq:Pe_A00_F}, and \eqref{eq:Pe_A01_F}  into \eqref{eq:Total_Prob_Generic}, the \gls{ber} conditioned on unsuccessful \gls{sic} can be expressed as
\begin{equation}
P_{\mathrm{e}|\mathcal{F}}^{\mathrm{opt}}=\frac{I_{11}^{\mathcal{F}}+I_{10}^{%
\mathcal{F}}}{q_{11}+q_{10}}.  \label{eq:Pe_F_final}
\end{equation}
\subsection{The Total  BER of the Optimal Detector}
The average \gls{ber} of the near-user using the optimal detector is 
\begin{equation}
\bar{P}_{\mathrm{e},2}^{\mathrm{opt}}=P_{\mathrm{e}|\mathcal{S}}^{\mathrm{opt}}\Pr
(\mathcal{S})+P_{\mathrm{e}|\mathcal{F}}^{\mathrm{opt}}\Pr (\mathcal{F}).
\label{eq:Pe_total_start}
\end{equation}%
Using \eqref{eq:Pr_SF}, \eqref{eq:Pe_S_final}, and \eqref{eq:Pe_F_final}, we
obtain 
\begin{equation}
{\bar{P}_{\mathrm{e},2}^{\mathrm{opt}}=\frac{1}{2}\left( I_{11}^{\mathcal{S}%
}+I_{10}^{\mathcal{S}}+I_{11}^{\mathcal{F}}+I_{10}^{\mathcal{F}}\right) }.
\label{eq:Pe_total_opt}
\end{equation}%
The cancellation of $p_{11}$, $p_{10}$, $q_{11}$, and $q_{10}$ in \eqref{eq:Pe_total_opt} does not imply that the post-\gls{sic} fading and
noise were treated as independent. These terms cancel only after the
conditional \glspl{ber} being weighted by the corresponding conditional priors
and \gls{sic} outcome probabilities. The dependence between fading and noise
is already accounted for through the joint post-\gls{sic} \glspl{pdf} in the
individual \gls{sic} outcome  integrations. The total \gls{ber} in \eqref{eq:Pe_total_opt} does not result in closed-form solution, however,  generally it requires one-dimensional numerical integration. 
\subsection{Comparison With the Conventional SIC/JML Detector}
According to \eqref{eq-JMLD-NU},  the conventional \gls{sic}/\gls{jml} detector sets $\tau_\beta = \beta \sqrt{\alpha _{1}}$. Consequently, substituting it into the four optimal-detector integrals in \eqref{eq:Pe_A00_S}, \eqref{eq:Pe_A01_S}, \eqref{eq:Pe_A00_F}, and \eqref{eq:Pe_A01_F}, gives $I_{11}^{\mathcal{S},\mathrm{SIC}} =G(A_{11})-G(\sqrt{\alpha _{2}})$, $I_{10}^{\mathcal{S},\mathrm{SIC}} =1-G(\sqrt{\alpha _{2}})$, $I_{11}^{\mathcal{F},\mathrm{SIC}} =1-G(2\sqrt{\alpha _{1}}+\sqrt{\alpha _{2}%
})$, and $I_{10}^{\mathcal{F},\mathrm{SIC}} =G(2\sqrt{\alpha _{1}}-\sqrt{\alpha _{2}}%
)-G(A_{10})$, respectively. Therefore, 
\begin{multline}
{\bar{P}_{\mathrm{e},2}^{\mathrm{SIC}}=}\frac{1}{4}\hspace{-1mm}\left[ 
2+\mathcal{M}\left(A_{11}\right)-\mathcal{M}\left(A_{10}\right)+\mathcal{M}\left(2\sqrt{\alpha _{1}}-%
\sqrt{\alpha _{2}}\right) \right. \\ 
\left.-\mathcal{M}\left(2\sqrt{\alpha _{1}}+\sqrt{\alpha _{2}}\right)-2\mathcal{M}\left(\sqrt{%
\alpha _{2}}\right)
\right]  \label{eq:Pe_SIC_closed}.
\end{multline}
\section{Numerical Results}
This section evaluates the \gls{ber} performance of the near-user using the optimal detector in \eqref{eq-ML-NU} and the conventional \gls{sic}/\gls{jml} detector in \eqref{eq-SIC-NU} and \eqref{eq-JMLD-NU}. The theoretical \gls{ber} results are obtained using \eqref{eq:Pe_total_opt}--\eqref{eq:Pe_SIC_closed}. The fading amplitude is Rayleigh distributed with $\Omega_k=1$, and the transmitted symbols of both users are equiprobable \gls{bpsk} symbols. Moreover, define $\bar{\gamma}=1/{\sigma_n^2}$.

\begin{figure}[tb]%
\centering
\includegraphics[
height=3.1176in,
width=3.1176in
]{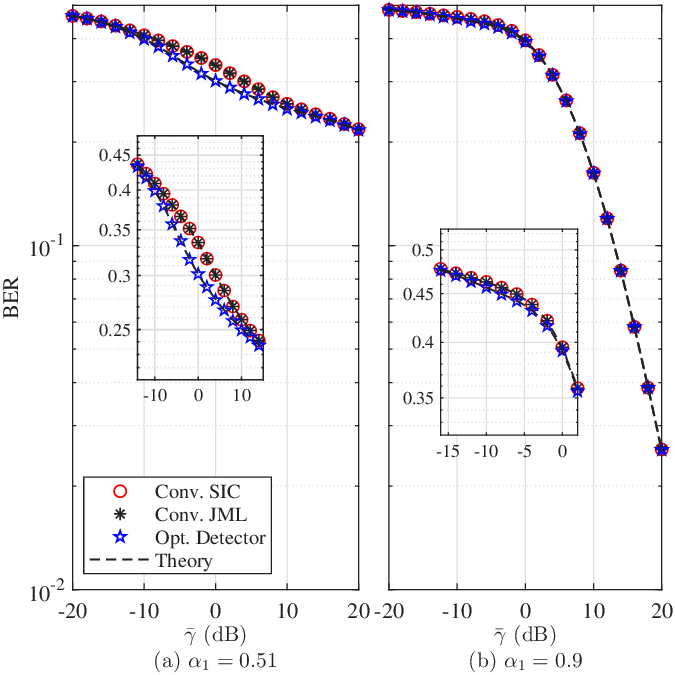}
\caption{The simulated and theoretical average \glspl{ber} of the conventional \gls{sic}, \gls{jml}, and optimal detector of $U_2$ with
(a) $\alpha_1=0.51$ and (b) $\alpha_1=0.9$.}
\label{Fig:Figure_1}
\end{figure}

Fig. \ref{Fig:Figure_1} compares the simulated and theoretical \gls{ber} of $U_2$ for (a) $\alpha_{1}=0.51$ and (b) $\alpha_{1}=0.9$. The simulation results match the analytical curves perfectly, validating the derived \gls{ber} expressions. The conventional \gls{sic} and \gls{jml} curves overlap, confirming that the two detectors induce identical near-user decision regions, in agreement with Remark \ref{remark4}. For $\alpha_{1}=0.51$, the optimal detector provides a moderate gain over the conventional \gls{sic}/\gls{jml} detector. This is due to the nearly balanced power allocation, and hence, the two Gaussian likelihoods overlap significantly and the \gls{map} boundary $\tau_{\beta}$ deviates from the conventional boundary $\beta\sqrt{\alpha_{1}}$. For $\alpha_{1}=0.9$, the far-user's signal dominates, thus the \gls{sic} decision becomes highly reliable, and the optimal and conventional detectors yield nearly identical \gls{ber}.

\begin{figure}[tb]%
\centering
\includegraphics[
height=3.1176in,
width=3.1176in
]{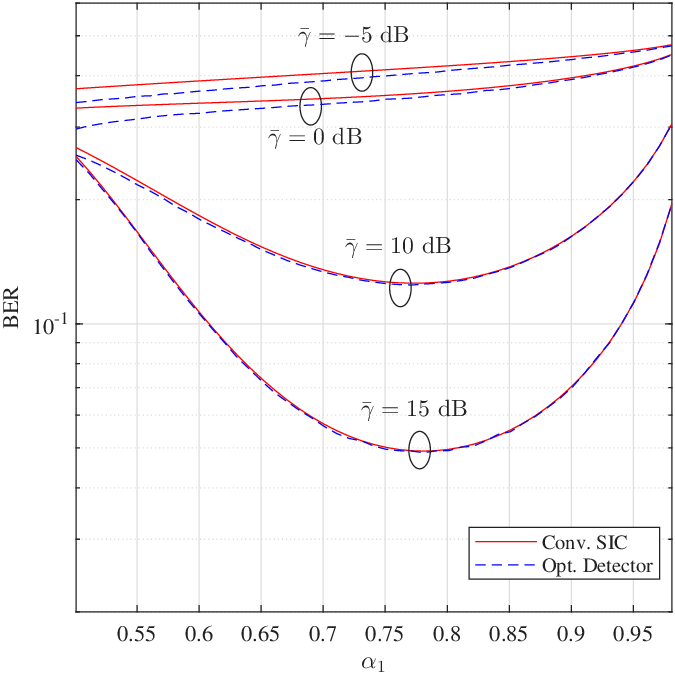}
\caption{The average \gls{ber} of the conventional \gls{sic} and optimal detectors versus $\alpha_1$, where $\alpha_1\in[0.51,0.99]$ and
$\bar{\gamma}\in\{-5,0,10,15\}$ dB.}
\label{Fig:Figure_2}
\end{figure}

Fig. \ref{Fig:Figure_2} shows the \gls{ber} of  $U_2$ as a function of $\alpha_{1}$ for selected values of $\bar{\gamma}$. At low \glspl{snr} such as $\bar{\gamma}=-5$ dB and $\bar{\gamma}=0$ dB, the \gls{ber} is noise dominated and varies slowly with $\alpha_{1}$. As $\bar{\gamma}$ increases, the \gls{ber} becomes more sensitive to the power allocation and exhibits a minimum at an intermediate value of $\alpha_{1}$, reflecting the fundamental \gls{noma} tradeoff where allocating more power to the far-user improves the reliability of the \gls{sic} stage but reduces the energy available for $s_{2}$. Across the entire range of $\alpha_{1}$ and \glspl{snr} considered, the optimal detector consistently achieves a lower \gls{ber} than the conventional \gls{sic}, with the largest improvement occurring at the low \gls{snr} and $\alpha_{1}$ regions, where $\tau_{\beta}$ deviates most from $\beta\sqrt{\alpha_{1}}$.

\section{Conclusion}
This letter derived the optimal detector for the two-user \gls{noma} system based on the \gls{map} criterion to minimize the individual users' \glspl{ber}. In addition, the condition under which the derived near-user detector reduces to the conventional \gls{sic} detector is identified, and the equivalence between conventional \gls{sic} and \gls{jml} detection is formally established. Furthermore, an exact \gls{ber} analysis accounting for the statistical impact of the \gls{sic} decision is presented for the derived detectors over Rayleigh fading channels. Extensions to higher-order modulation schemes and an arbitrary number of users are left for future work. 
\vspace{-2mm}
\bibliographystyle{IEEEtran}

\end{document}